\documentclass[twocolumn,showpacs,prl]{revtex4}
\usepackage{graphicx}
\usepackage{bbold}
\usepackage{amsmath}
\usepackage{epsfig}
\usepackage{subfigure}
\usepackage{sidecap}
\usepackage{setspace}

\newcommand{\eqa}{\begin{eqnarray}}
\newcommand{\eqe}{\end{eqnarray}}

\newcommand{\bm}{{\bf m}}

\newcommand{\bu}{{\bf  u}}
\newcommand{\bx}{{\bf x}}

\newcommand{\br}{{\bf r}}

\begin{document}

\title{Current Induced Vortex Wall Dynamics in Helical Magnetic Systems}

\author{Bahman Roostaei}
\affiliation {Department of Physics, Indiana University-Purdue University Indianapolis, Indiana 46202, USA}
\affiliation {Department of Physics, Institute For Research In Fundamental Sciences, Tehran, Iran.}
\date{\today}
\begin{abstract}
Nontrivial topology of interfaces separating phases with opposite chirality in helical magnetic metals result in new effects as they interact with spin polarized current. These interfaces or vortex walls consist of a one dimensional array of vortex lines. We predict that adiabatic transfer of angular momentum between vortex array and spin polarized current will result in topological Hall effect in multi-domain samples. Also we predict that the motion of the vortex array will result in a new damping mechanism for magnetic moments based on Lenz's law. We study the dynamics of these walls interacting with electric current and use fundamental electromagnetic laws to quantify those predictions. On the other hand discrete nature of vortex walls affects their pinning and result in low depinning current density. We predict the value of this current using collective pinnng theory.
\end{abstract}
{}

\maketitle
 Localized magnetic moments in some rare earth metals such as Ho, Tb, Dy\cite{Jensen, Lang2004} and their alloys, interact through anisotropic RKKY exchange interaction. This interaction consists of ferromagnetic nearest neighbor and antiferromagnetic next nearest neighbor between planes while the in-plane interactions are ferromagnetic\cite{Jensen}. As the result for example in Ho at temperatures below 133 K a helical order is developed with the helical axis perpendicular to the basal planes and below 19 K the moments cant away from the planes to form a conical phase\cite{Lang2004}.

The ground state of this helical system is degenerate with respect to the sense of rotation of the helices or chirality of the structure. Recently observation of chiral domains in these systems\cite{Lang2004} has triggered novel ideas and predictions about the interface between chiral domains in a general perspective\cite{Li2012,Nattermann,Roostaei}. The most striking one is the prediction that these interfaces posses vorticity totally unrelated to the boundary effects contrary to what is seen for example in the case of vortex domain walls (consisting of a single vortex) in certain ferromagnetic nanowires\cite{Nakatani}. More precisely these interfaces consist of vortex lines parallel to the basal planes.

Early studies have shown that the peculiar structure of these walls leads to a different static and dynamic behavior. The interaction between the vortex lines have been predicted to be weak\cite{Roostaei} which implies the elasticity must originate from a vicinal behavior\cite{Nattermann}. On the other hand the movement of the vortex lines inside the plane of the interface leads to the mobility enhancement\cite{Roostaei}.

Here in the present work we focus on the interaction of the spin polarized current with these walls. The vorticity of the walls rotates the conduction electron's spins. This torque can be seen by electrons as a fictitious magnetic field generated by the wall. The interplay between this induced \textit{Berry phase} and electron motion leads to topological Hall effect as the first result in this letter. This is an \textit{anomaly in resistivity caused by domain walls}.

The next result is a dominating mechanism for damping of the local magnetic moments as a result of vorticity and the domain wall motion. Local fluctuations and motions of the wall will add time dependent phase to the wavefunction of the conduction electrons which can be interpreted as an extra electric field (the so called spin motive force) which at the end drags the motion of domain wall. We will show that this dynamical effect will generate an additional damping of the magnetic moments.

On the other hand the the angular momentum transfer from conduction electrons to the domain wall results in a force which can move the domain wall. We predict in the presence of weak disorder the current necessary to depinn the wall is smaller than predicted values for walls between ferromagnetic domains, a promising property for future technological applications\cite{Parkin2008}.

The discretized Hamiltonian describing the helical magnetic order in the rare earth metals we are considering is:
\eqa
 H&=&-J\sum_{\langle ij\rangle, n} {\bf m}_{i,n}\cdot {\bf m}_{j,n}-J\sum_{i,n}{\bf m}_{i,n}\cdot {\bf m}_{i,n+1}+\nonumber\\
 &+& J'\sum_{i,n}{\bf m}_{i,n}\cdot {\bf m}_{i,n+2}
\eqe
which introduces a ferromagnetic(antiferromagnetic) exchange coupling $J>0$ $(J'>0)$ between local moments ${\bf m}_{i,n}$ in basal plane number $n$. We parameterize the local magnetic moments by spherical coordinates $\bm=\{\sin\theta\cos\varphi,\sin\theta\sin\varphi,\cos\theta\}$ so we can write the long range Hamiltonian of the magnetic system as:
\eqa
H_S\approx{J\over a}\int_\br \left\{\left(\nabla_\perp\varphi\right)^2+{a^2\over 4}\left[\left(\partial_z\varphi\right)^2-q^2\right]^2+{a^2\over 4}\left(\partial_z^2\varphi\right)^2\right\}.\nonumber\\
\eqe
We consider the polar angle $\theta$ being almost constant throughout the system. In the above the helix propagation wavevector $q=\Theta/a=a^{-1}\arccos(J/4J')$ and $\Theta$ is the angle which moments in successive planes make with respect to each other. $a$ is the lattice constant of the crystal and we assume it is the same for all directions. Also we have taken z-axis perpendicular to the basal planes and $\nabla_\perp=\{\partial_x,\partial_y\}$.

\begin{figure}
\includegraphics[scale=0.3]{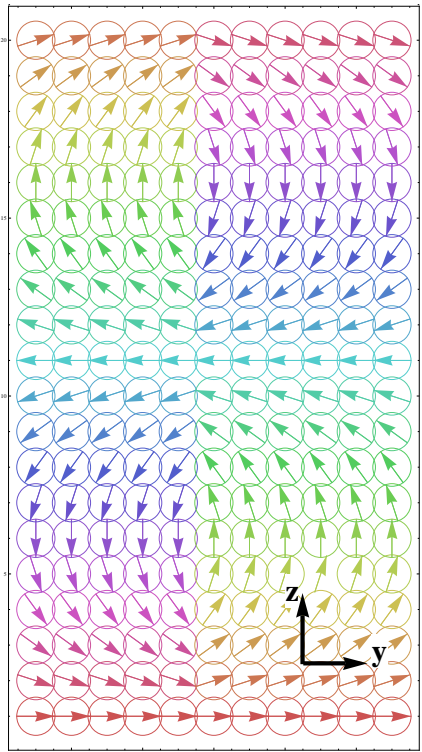}
\caption{Illustration of vorticity. The right and left helix have opposite chirality. For better visualization spins have been represented to lie in the $y-z$ plane, \textit{parallel} to the helical axis while in reality they lie in $x-y$ plane.}
\label{path_integral}
\end{figure}

Obviously the above Hamiltonian is degenerate with respect to two different chiral ground states $\varphi=\pm q z$. Li \textit{et al.}\cite{Li2012} have realized the domain wall between chiral states must consist of an array of vortex lines. This can be easily seen by performing the line integral $\oint \nabla\varphi.d{\bf l}$ on a path encircling any part of the interface (in x-z plane in figure \ref{path_integral}). So from now on by vortex wall we mean a one dimensional array of vortex lines with a rest separation of $\pi/q$ ($\approx 2.1 nm$ for Ho) which is half the length of one helix.

Conduction electrons moving inside a magnetic environment interact through Hund's rule with local magnetic moments at each point. In an adiabatic approximation we can assume the local magnetization acts as a local magnetic field and aligns the spin of electrons with its direction. This adiabatic approximation is satisfied when the typical variation of the local moment direction is much larger than the electron wavelength, $k_F\gg q$ which is held in most rare earth metals\cite{Jensen}. With this assumption one can use the well known adiabatic approximate for the electrons by projecting their spins into the subspace of the local magnetic structure. In this approximation the wavefunction of electron can be written as $|\psi\rangle=\phi|{\bf m}\rangle$ in which $\phi$ is the orbital and $|\bm\rangle$ is the spin part of the wave function, ${\boldsymbol\sigma}\cdot\bm|\bm\rangle=\bm|\bm\rangle$ matching the local magnetic distribution of the metal.

With this map, the Hamiltonian of a single electron $\hat{H}={{\bf p}^2\over 2m}+V(\br)+J_H {\boldsymbol\sigma}\cdot{\bf m}$ will look like the Hamiltonian of an electron in presence of an external electromagnetic field:
\eqa
\hat{H}_{eff}\approx {1\over 2m}({\bf p}-e{\bf a})^2-ea_0 + V(\br)
\eqe
in which the fictitious vector potential is $a_\mu={\hbar\over 2e}(1-\cos\theta)\partial_\mu\varphi$\cite{Tatara08}. In this subspace the spin part of the Hamiltonian is now diagonalized and acts as a constant Zeeman term with a coupling proportional to Hund's coupling $J_H$. This term is approximately constant and so ineffective for slowly varying textures in all of our calculations.

In the simplest case of a flat vortex domain wall electrons that are moving transverse to the wall plane with a velocity $v$ with respect to the wall, feel the fictitious vector potential ${\bf a}\approx {\hbar\over 2e}\nabla\varphi$ (assuming $\theta\approx \pi/2$) which is corresponding to a magnetic field ${\bf B}=\nabla\times {\bf a}={\hbar\over 2e}\nabla\times\nabla\varphi={h\over 2e}\sum_n \delta[{\bf x}_\perp-{\bf x}_n(t)]{\hat{\bf x}}$ in which ${\bf x}_\perp=\{y,z\}$ and ${\bf x}_n(t)=\{-vt,n\pi/q\}$ represents a vortex line.

The magnetic field that electrons feel is a direct consequence of the vorticity of the wall and deflects electron's path resulting in a \textit{topological Hall} contribution, $\sigma_{yz}^H$ to the conductivity of the metal which can be estimated using:
\eqa
\sigma_{yz}^H/\sigma_{yy}\approx e B_x \tau/m
\eqe
The magnetic flux density here corresponds to one vortex line $B_x\approx \phi_0/(\pi/q)^2$ and $\tau$ is the typical scattering time of the metal.

This topological Hall conductivity is the result of the exerted torque on the spin of electron by the vortex domain wall and in principle will cause anomaly in magnetotransport in multi-domain samples for example in Ho below 133 K for which with typical values of scattering time $\sigma_{yz}^H/\sigma_{yy}\approx 0.1$. This effect has been already realized in skyrmion lattices in some noncentrosymmetric magnetic systems such as MnSi and also in magnetic nanowires and disks\cite{Neubauer, Hall}.

\textit{Wall equation of motion}: We parameterize each vortex line in the vortex domain wall by the collective coordinates ${\bf u}(x,n,t)=\left\{u_y(x,n,t),n\pi/q+u_z(x,n,t)\right\}$ with respect to the wall's rest position. We assume the vortices are rigid and consider long wavelength distortions of the domain wall $|{\bf u}(x,n,t)-{\bf u}(x',m,t)|^2\ll (n-m)^2\pi^2/q^2+(x-x')^2$ and build a long wavelength Lagrangian for this texture.

The kinetic energy part of the Lagrangian reflects the spin dynamics or the Berry phase of each spin:
\eqa
K={\hbar S\over a^3}\int d^3{\br} (\cos\theta-1)\dot{\varphi}
\eqe
This energy in terms of the vortex coordinates will be $K\approx {q\over 2\pi a}\int d^2\bx_\parallel {\bf G}\cdot[\dot{\bu}\times \bu ]$. Here $\bu(\bx_\parallel,t)$ is the coordinate of each vortex line in continuum limit and $\bx_\parallel=\{x,z\}$. The average Gyrovector of the vortex line is introduced also as ${\bf G}={\hbar S\over a^3}(\pi/q)p{\hat{\bf x}}$ where $p$ is the average dimensionless value of local moment component out of the basal plane direction. This value is nonzero in conical phase of Holmium.

Gyrovector is always perpendicular to the plane of the vortex. In most of the conventional systems the moments forming the vortex lie inside the plane of the vortex however in our case the moments lie inside the $x-z$ plane instead of $x-y$ plane. This is why the Gyrovector is not perpendicular to the plane of the vortex: Indeed Gyrovector is a second rank tensor, $G_{ij}=\int d^2r {\bf m}\cdot\partial_i {\bf m}\times \partial_j {\bf m}$. In our system the out of plane component ($\cos\theta(y)$) of moments vary slowly as a function of $y$ as one approaches from $y\rightarrow \infty$ toward the center of the wall $y=0$. Assuming full symmetry in the $x$-direction the only nonzero components of this tensor are $G_{yz}=-G_{zy}$. This leads to a pseudovector in the $x$-direction.

The interaction of the spin current with the domain wall is also obtained by summing over contribution from all vortex lines:
\eqa
 H_{int}=-\int d^3\br {\bf j}\cdot {\bf a}\approx {q\over\pi a}\int d^2\bx_\parallel {\bf G}\cdot[{\bf v}_s\times \bu]
\eqe
Here we have defined electron's velocity ${\bf v}_s=a^3/(2eS){\bf j}$ considering perfect polarization of the current. Finally the energy cost for distortions of the domain wall can be approximated by a local elastic term\cite{Li2012,Nattermann}:
\eqa
H_{el}={\sigma\over 2}\int d^2\bx_\parallel |\nabla\bu|^2.
\eqe
Damping of the local moments can be incorporated into the equation of motion by adding the dissipation function to the Lagrange equation\cite{Gilbert}:
\eqa
W=\hbar S\alpha_d/2a^3\int d^3 \dot{\bm}^2={q\alpha_d\over 2\pi a}D\int d^2\bx_\parallel \dot{\bu}^2
\eqe
Here $\alpha_d$ is the Gilbert damping constant and $D={\hbar S\over a^3}\pi/q\int dxdy |\nabla \bm_0|^2$ is the profile function for one vortex which behaves as logarithm of the ratio of the system size to the vortex core size.

We can now construct the total Lagrangian of the domain wall $L=K-H_{int}-H_{el}$ and obtain its equation of motion\cite{Shibata}:
\eqa\label{eq.motion}
\dot{\bf u}={\pi a\sigma\over qG^2}{\bf G}\times\nabla^2{\bf u}+{\bf v}_s+{\alpha_d D\over G^2}\dot{\bu}\times {\bf G}
\eqe
Note that the dissipation function enters the Lagrange equation by derivative of $W$ with respect to $\dot{\bu}$.

The first conclusion is that a flat vortex wall will start moving with the same velocity as electrons in the absence of damping. On the other hand the boundaries of the sample acts as an effective potential to keep the vortex line array inside the system. Damping torque will distort the vortex lines in transverse and longitudinal directions since $u_z$ and $u_y$ are conjugate of each other\cite{nagaosa}. However the average longitudinal movements $u_z$ will be confined by the boundary.

\textit{Damping}: As the result of the vortex wall movement the conduction electrons feel a time dependent magnetic flux density. This magnetic flux density vector, assuming symmetry in the $\hat{\bf z}$ direction is:
\begin{equation}
{\bf B}({\bf r},t)=\phi_0\rho_v({\bf \bx_\perp},t){\hat{\bf x}}
\end{equation}
in which $\rho_v(\bx_\perp,t)=\sum_n\delta^2[\bx_\perp-{\bf u}_n(t)]$ is the vortex wall vortex density and ${\bf u}_n(t)$ represents an undistorted vortex line. Here we ignore the distortions of the vortex wall.

The resulting change in the magnetic flux according to Lenz's law creates an electric field which in the conduction electron's rest frame will induce another current depending on the conductivity of the medium. Let's calculate this electric field. The time derivative of the magnetic flux density is:
\begin{equation}
\nabla\times {\bf E} = -{\partial {\bf B}\over \partial t} = -\phi_0 {\partial\rho_v\over \partial t}{\hat{\bf z}}
\end{equation}
Since the total charge of the system is zero we have $\nabla\cdot{\bf E} = 0$. The solution is of the form ${\bf E}={\bf E}_\perp\times {\bf\hat{x}}$ satisfying $\nabla\cdot{\bf E}_\perp=\phi_0{\partial\rho_v\over \partial t}$. We integrate this equation on a cylinder with radius $\approx \pi/q$ around each vortex line to find ${\bf E}_\perp$. The \textit{induced rotating electric field}, $\bf E$ will push the electrons and create a current ${\bf j}_d=\sigma_0{\bf E}$ in which $\sigma_0$ is the conductivity of the medium. The resulting contribution in conduction electron velocity will be:
\begin{equation}
{\bf v}_d=\alpha_d'\dot{{\bf u}}\times {\bf\hat x}
\end{equation}
where we have defined the new damping constant:
\eqa
\alpha_d'={\phi_0}{a^3\sigma_0\over 4\pi eS(\pi/q)^2}
\eqe

This will in turn induce a \textit{drag} force on the domain wall and contribute to the damping of its motion. The ratio of the damping terms in the equation of motion (\ref{eq.motion}) of the domain wall will be then:
\begin{equation}
r_d=\phi_0{1\over (\pi/q)^2f}{\sigma_0\over \alpha_d n_s}
\end{equation}
in which $n_s=2eS/a^3$ the electron spin density, $f=\int_{\bf r}|\nabla{\bf m}_0|^2$ is the vortex profile function introduced before. To estimate this ratio we use the typical values for Holmium\cite{Jensen,Resistance}, $J\approx 0.7 meV$, $\Theta\approx \pi/6$, $\sigma_0^{-1}\approx 60 \mu\Omega.cm$, $\alpha_d\approx 0.005$ and sample size of a micron which result in a value $r_d\approx 6.7$. Thus this effect can be dominating in principle.

The emergent damping mechanism introduced above is originated from the time dependence of the vector potential ${\bf a[\dot{\bm}}]$ which has lead to generation of an electric current. This current affects the dynamics of moments through the term ${\bf j}_d\cdot\nabla \bm$ in Landau-Lifshitz-Gilbert equation. This in turn has lead to appearance of the damping term in the equation of motion of the domain wall thus maybe distinguished in FMR linewidth measurements as domains are removed. Also note that the origin of Gilbert damping ($\alpha_d$) is the spin-orbit coupling which eventually relaxes the magnetic moments and violates the conservation of total spin. On the other hand this new damping term in equation of motion of the domain wall vanishes as the system approaches uniform (as opposed to non-collinear) magnetization ($\alpha_d'\rightarrow 0 $ as $ q\rightarrow 0 $).

\textit{Depinning from Disorder}: There are two types of disorder that can pin the vortex domain wall. The one we discuss is the vacancies or non-magnetic impurities. The vortex lines in the wall could gain exchange energy by moving into those points. The second type is the charged impurities which cause some local variation in electron density hence variation in exchange energy $J$. This will create a random potential that can collectively pin the wall. Here in this study we consider only vacancy disorder.

We assume vacancies or non-magnetic impurities are spheres occupying lattice sites in random with radius $a$ of the order of lattice constant. Then the total wall disorder energy will be:
\begin{equation}
H_{ni}=\int d^3r \sum_{i=1}^{N_p}\sum_n a^3\varepsilon_v\left[\bx_\perp-\bu(x,n)\right]\delta_a^3({\bf r}-{\bf r}_i)
\end{equation}
in which $\varepsilon_v(\bx_\perp)$ is the energy per unit volume of a straight vortex line, ${\bf r}_i$ is the vacancy position and $\delta_a(x)$ is a smooth delta function on a scale $a$. This energy can be written in continuum limit as $H_{ni}\approx\int d^2{\bf x}_\parallel V_D({\bf x}_\parallel,u_y)\rho(z-u_z)$ where $\rho(z)=\sum_n\delta(z-n\pi/q)$. The average of random potential energy density $V_D(\br)$ vanishes and its correlation is:
\begin{equation}
\Delta({\bf r})=a^6n_i\delta_a(x)\int d^2\bx_\perp' \varepsilon_v(\bx_\perp-\bx_\perp')\varepsilon_v(\bx_\perp')
\end{equation}
in which $n_i$ is the impurity density. By average we mean averaging over all equally probable sets of $\{\br_i\}$.

The disorder potential exerts torque on the vortex lines. This torque will be added to the equation of motion (\ref{eq.motion}) of the vortex wall via the term $-{\pi a\over q G^2}{\bf G}\times \nabla_{\bf u}\left[V({\bf x}_\parallel,u_y)\rho(z-u_z)\right]$. Let's estimate the value of the current density necessary to depinn the vortex wall in the weak pinning regime. The typical variations of the disorder energy for a set of $N$ vortex lines $(N\gg 1)$ each with a length $L$ is:
\begin{equation}
\left\langle H_{ni}^2\right\rangle \approx  {q\over\pi}LN{\tilde{\Delta}(0)\over \xi_d}
\end{equation}
where tilde sign mean Fourier transform. We are assuming the correlations of the disorder distribution is of short range $(\xi_d)$. The average elastic energy of this part of the array assuming they have distorted by $\pi/q$ is $\sigma \pi^2/q^2$. Thus the total energy cost per vortex line per unit of length will be:
\eqa
\mathcal{E}\approx \sigma \pi^2/N L q^2-[{q{\tilde{\Delta}(0)}/\pi\xi_d}]^{1/2}/\sqrt{NL}.
\eqe
Minimizing this eergy will give the so called Larkin length\cite{Blatter} ${\cal L}=NL\approx 4\pi^5\sigma^2\xi_d/q^5\tilde{\Delta}(0)$.

Depinning threshold is when the external torque is compensated by the disorder torque at Larkin length scale:
\begin{equation}
j_c\approx {2eS\over a^3}{\pi a\over q G^2}\left\langle \left|{\bf G}\times \nabla_{\bf u}\left[V({\bf
x}_\parallel,u_y)\rho(z-u_z)\right]\right|\right\rangle_{\cal L}
\end{equation}
The disorder energy density at the Larkin length scale per vortex line per unit length is $\left \langle V\rho\right\rangle_{\cal L} \approx q^4\tilde{\Delta}(0)/2\pi^4\xi_d\sigma$. From this energy density we can find out the typical disorder torque at Larkin length assuming the domain wall distortion is of the order of $\pi/q$ in this scale: $\left\langle|\nabla_{\bf u}(V\rho)|\right\rangle \approx q^5\tilde{\Delta}(0)/2\pi^5\xi_d\sigma $. This estimation gives us an estimate for the threshold current density:
\begin{equation}
j_c\approx {eS/a^2\over (\pi/q)^4|{\bf G}|}{\tilde{\Delta}(0)\over \sigma\xi_d}
\end{equation}
Because of strong screening of the vortices by helical background (the vortex effect on local magnetic moments vanishes very fast away from the wall into the helical phase)\cite{Roostaei} the vortex energy density vanishes at distances longer than $\pi/q$ consequently we can use an approximate exponential model for  the energy density of vortex lines and find the above correlation function:
\begin{equation}
{\Delta}({\bf r})\approx {\sigma^2\over q^2}a^6n_i\delta_a(x)\delta_{\pi/q}^2(\bx_\perp)
\end{equation}
where we have used the fact that the total energy of one vortex line is $\approx \sigma L\pi/q$. From this we will have $\tilde{\Delta}(0)\approx \sigma^2n_ia^6\pi^2/q^2$ and:
\begin{equation}
j_c \approx {1\over 2\pi^4}{e\over \hbar}(n_ia^3){(aq)^3\over p}(\sigma a^2)\left({1\over a\xi_d}\right)
\end{equation}

The energy of one vortex line has been calculated easlwhere\cite{Li2012} using variational calculation which results in $\sigma\approx Jq/a$. Also we approximate $\xi_d\approx \pi/q$. Again for Holmium and typical disorder concentration $n_i a^3\approx 10^{-5}$ and $p\approx 0.1$ (conical phase) we obtain $j_c\approx 5\times 10^5$  A.m$^{-2}$ which is much lower than the value observed for more conventional domain walls\cite{Yamaguchi}.

In conclusion, we have presented a microscopic theory for the interaction of spin polarized current with a new type of magnetic domain wall. This theory predicts the existence of a topological Hall effect because of the topological nature of the wall. Also the collected Berry phase by the electron motion interacting with a moving vortex wall affects their motion which in turn damps the wall motion. One should note that besides the Gilbert damping term there is another higher order term also associated with the relaxation of spins, the so called $\beta$-term\cite{Tatara08}. This term is almost comparable (if not smaller)\cite{Shibata2011} in magnetic metals to the Gilbert damping term. That is why it is plausible to assume the drag force found in this work to be an effective force needed to be considered in vortex domain wall dynamics under spin polarized current.

Finally this current can depinn the wall from non-magnetic impurity sites which have collectively pinned the wall. Present theory predicts a very low depinning current which together with the lack of intrinsic pinning for vortices makes the vortex walls in helical magnets attractive for technological applications.

I would like to thank Dr. Yogesh N. Joglekar and Department of Physics at Indiana University - Purdue University Indianapolis for supporting this work.

\end{document}